\documentclass[aps,pra,preprint,groupedaddress, superscriptaddress, citesort]{revtex4}

\usepackage{amssymb,}
\usepackage{amsmath}
\usepackage{graphics}
\usepackage{graphicx}
\usepackage{float}
\def\be{\begin{equation}}
\def\ee{\end{equation}}
\def\ba{\begin{array}}
\def\ea{\end{array}}

\begin{document}

\title{A Note on Holevo quantity of $SU(2)$-invariant states}
\author{Yao-Kun Wang}
\affiliation{College of Mathematics,  Tonghua Normal University, Tonghua, Jilin 134001, China}
\author{Li-Zhu Ge}
\affiliation{The Branch Campus of Tonghua Normal University, Tonghua, Jilin 134001, China}
\author{Shao-Ming Fei}
\affiliation{School of Mathematical Sciences,  Capital Normal University,  Beijing 100048,  China}
\affiliation{Max-Planck-Institute for Mathematics in the Sciences, 04103 Leipzig, Germany}
\author{Zhi-Xi Wang}
\affiliation{School of Mathematical Sciences,  Capital Normal
University,  Beijing 100048,  China}

\begin{abstract}
The Holevo quantity and the $SU(2)$-invariant states have particular importance in quantum information processing. We calculate analytically the Holevo quantity for bipartite systems composed of spin-$j$ and spin-$\frac{1}{2}$ subsystems with $SU(2)$ symmetry, when the projective measurements are performed on the spin-$\frac{1}{2}$ subsystem. The relations among the Holevo quantity, the maximal values of the Holevo quantity and the states are analyzed in detail. In particular, we show that the Holevo quantity increases in the parameter region $F<F_d$ and decreases in region $F>F_d$ when $j$ increases, where $F$ is function of temperature in thermal equilibrium and $F_d=j/(2j+1)$, and the maximum value of the Holevo quantity is attained at $F=1$ for all $j$. Moreover, when the dimension of system increases, the maximal value of the Holevo quantity decreases.
\end{abstract}
\maketitle

\section{Introduction}

The Holevo bound characterizes the capacity of quantum states in classical communication \cite{holevo,bena}. As an exceedingly useful upper bound on the accessible information it plays an important role in many quantum information processing \cite{nielsen}.
It is a keystone in the proof of many results in quantum information theory \cite{lupo,zhang,lloyd,roga,wu,guo}. With respect to the Holevo bound,
the maximal Holevo quantity related to weak measurements has been studied in \cite{wang}.
The Holevo quantity of an ensemble $\{p_{i};\rho_{A|i}\}$, corresponding to a bipartite quantum state $\rho_{AB}$ with the projective measurements $\{\Pi^B_i\}$ performed on the subsystem $B$, is given by \cite{wang}
\begin{eqnarray}\label{fo}
\chi\{\rho_{AB}|\{\Pi^B_i\}\}=\chi\{p_{i};\rho_{A|i}\}\equiv S(\sum_{i}p_{i}\rho_{A|i})-\sum_{i}p_{i}S(\rho_{A|i}),
\end{eqnarray}
where
\begin{equation}\label{post}
p_i = \mbox{tr}_{AB}[(I_A \otimes \Pi^B_i ) \rho_{AB} ( {I}_A \otimes \Pi^B_i) ],~~
\rho_{A|i} = \frac{1}{p_i} \mbox{tr}_B[({I}_A \otimes \Pi^B_i) \rho_{AB} ({I}_A \otimes \Pi^B_i)].
\end{equation}
It denotes the A's accessible information about the B's measurement outcome when B projects its system by the projection operators $\{\Pi^B_i\}$.

When measuring one particle in a composite quantum system, partial measurement in quantum mechanics has recently been generalized in \cite{long} . It collapses-in or collapses-out the measured states. Measure of correlation is also very important in quantum gravity \cite{gyongyosi,van}, and characterizing coherence in quantum systems \cite{han}.

The Holevo quantity (\ref{fo}) related to projective measurements depends on the bipartite states. In the following we study the Holevo quantity for a particular class of $SU(2)$-invariant states. The $SU(2)$-invariant states originate from thermal equilibrium states of spin systems with $SU(2)$-invariant Hamiltonian \cite{Durkin}. With respect to two spins $\vec S_{1}$ and $\vec S_{2}$, a density matrices $\rho$ is said to be $SU(2)$-invariant
if $U_1\otimes U_2\rho U^{\dagger}_1\otimes U^{\dagger}_2=\rho$, where $U_{a}=\exp(i\vec\eta\cdot\vec S_{a})$, $a\in\{1,2\}$, are the usual rotation operator representation of $SU(2)$ with real parameter $\vec\eta$ and $\hbar=1$ \cite{Schliemann1,Schliemann2}. Those $SU(2)$-invariant
states $\rho$ commute with the total spin $\vec J=\vec S_{1}+\vec S_{2}$.
The entanglement of SO(3)-invariant bipartite quantum states have been studied in
\cite{Breuer1,Breuer2}. For the $SU(2)$-invariant quantum
spin systems it is shown that the negativity is the necessary and
sufficient condition of separability \cite{Schliemann1,Schliemann2}.
The relative entropy of entanglement has been analytically
calculated in \cite{zwang} for $(2j+1)\otimes 2$ and $(2j+1)\otimes 3$
dimensional $SU(2)$-invariant quantum spin states. The entanglement of formation, I-concurrence, I-tangle and convex-roof-extended negativity of the $SU(2)$-invariant states
of spin-$j$ and spin-$\frac{1}{2}$ systems have been investigated in \cite{Manne} by using the approach in \cite{Vollbrecht}. The quantum discord and one way deficit for the $SU(2)$-invariant states in spin-$j$ and spin-$\frac{1}{2}$ bipartite systems have been discussed respectively in \cite{cakmak,wang1}.

\section{Holevo quantity of $SU(2)$-invariant states}

As an $SU(2)$-invariant
state commutes with all the components of $\vec J$, $\rho$ has the general form \cite{Schliemann1},
\begin{equation}
\rho=\sum_{J=|S_{1}-S_{2}|}^{S_{1}+S_{2}}
\frac{A(J)}{2J+1}\sum_{J^{z}=-J}^{J}
|J,J^{z}\rangle_{0}{_{0}\langle J,J^{z}|}\,,
\end{equation}
where the constants $A(J)\geq 0$, $\sum_{J}A(J)=1$,
$|J,J^{z}\rangle_{0}$ denotes the state of total spin $J$ and
$z$-component $J^{z}$. We discuss the case with $\vec S_{1}$ of arbitrary spin $j$ and
$\vec S_{2}$ of spin $\frac{1}{2}$. A general $SU(2)$-invariant density matrix has the form,
\begin{eqnarray}
\rho^{ab}=\frac{F}{2j}\sum_{m=-j+\frac{1}{2}}^{j-\frac{1}{2}}|j-\frac{1}{2}, m\rangle
\langle j-\frac{1}{2}, m|+\frac{1-F}{2(j+1)}\sum_{m=-j-\frac{1}{2}}^{j+\frac{1}{2}}|j
+\frac{1}{2}, m\rangle\langle j+\frac{1}{2}, m|, \label{1state}
\end{eqnarray}
where $F\in[0,1]$, which is a function of temperature in thermal
equilibrium. $\rho^{ab}$ is a $(2j+1)\otimes 2$ bipartite state. It has two eigenvalues
$\lambda_1={F}/{(2j)}$ and $\lambda_2=({1-F})/({2j+2})$ with
degeneracies $2j$ and $2j+2$, respectively \cite{cakmak}.
As the eigenstates of the total spin can be given by the
Clebsch-Gordon coefficients \cite{shankar} in coupling a spin-$j$
to spin-$\frac{1}{2}$,
\begin{eqnarray}
|j\pm\frac{1}{2},m\rangle=
\pm\sqrt{\frac{j+\frac{1}{2}\pm m}{2j+1}}
|j,m-\frac{1}{2}\rangle\otimes
|\frac{1}{2},\frac{1}{2}\rangle
+\sqrt{\frac{j+\frac{1}{2}\mp m}{2j+1}}
|j,m+\frac{1}{2}\rangle\otimes
|\frac{1}{2},-\frac{1}{2}\rangle,\nonumber
\end{eqnarray}
the density matrix (\ref{1state}) can be written in product basis form \cite{cakmak},
\begin{eqnarray}
\rho^{ab}&=&\frac{F}{2j}\sum_{m=-j+\frac{1}{2}}^{j-\frac{1}{2}}
(a_-^2|m-\frac{1}{2}\rangle\langle m-\frac{1}{2}|\otimes
|\frac{1}{2}\rangle\langle \frac{1}{2}| \\\nonumber
& &+a_-b_-(|m-\frac{1}{2}\rangle\langle m+\frac{1}{2}|\otimes |\frac{1}{2}\rangle\langle -\frac{1}{2}| \\\nonumber
& &+|m+\frac{1}{2}\rangle\langle m-\frac{1}{2}|\otimes |-\frac{1}{2}\rangle\langle \frac{1}{2}|) \\\nonumber
& &+b_-^2|m+\frac{1}{2}\rangle\langle m+\frac{1}{2}|\otimes |-\frac{1}{2}\rangle\langle -\frac{1}{2}|)\\\nonumber
& & +\frac{1-F}{2(j+1)}\sum_{m=-j-\frac{1}{2}}^{j+\frac{1}{2}}(a_+^2|m-\frac{1}{2}\rangle\langle m-\frac{1}{2}|\otimes |\frac{1}{2}\rangle\langle \frac{1}{2}| \\\nonumber
& &+a_+b_+(|m-\frac{1}{2}\rangle\langle m+\frac{1}{2}|\otimes |\frac{1}{2}\rangle\langle -\frac{1}{2}| \\\nonumber
& &+|m+\frac{1}{2}\rangle\langle m-\frac{1}{2}|\otimes |-\frac{1}{2}\rangle\langle \frac{1}{2}|) \\\nonumber
& &+b_+^2|m+\frac{1}{2}\rangle\langle m+\frac{1}{2}|\otimes |-\frac{1}{2}\rangle\langle -\frac{1}{2}|),\nonumber
\end{eqnarray}
where $a_{\pm}=\pm\sqrt{\frac{j+\frac{1}{2}\pm m}{2j+1}}$ and
$b_{\pm}=\sqrt{\frac{j+\frac{1}{2}\mp m}{2j+1}}$.

Any von Neumann measurement on the spin-$\frac{1}{2}$ subsystem can
be written as $B_k=V\Pi_kV^{\dag}$, $k=0,1,$
where $\Pi_k=|k\rangle\langle k|$, $|k\rangle$ is the computational basis,
$V=tI+i\vec{y}\cdot\vec{\sigma}\in SU(2)$, $\vec{\sigma}=(\sigma_1,\sigma_2,\sigma_3)$ with $\sigma_1,\sigma_2,\sigma_3$ being Pauli matrices,
$t\in \Re$ and $\vec{y}=(y_{1},y_{2},y_{3})\in \Re^{3}$, $t^2+y_1^2+y_2^2+y_3^2=1$.
After the measurement, $\rho^{ab}$ has the ensemble of
post-measurement states $\{\rho_k, p_k\}$, with the post-measurement states $\rho_k$ and the corresponding probabilities $p_k$,
\begin{eqnarray}
p_k\rho_k &=&(I\otimes B_k)\rho^{ab}(I\otimes B_k)=(I\otimes V\Pi_kV^{\dag})\rho^{ab}(I\otimes V\Pi_kV^{\dag}) \\ \nonumber
&=&(I\otimes V)(I\otimes \Pi_k)(I\otimes V^{\dag})\rho^{ab}(I\otimes V)(I\otimes \Pi_k)(I\otimes V^{\dag}). \nonumber
\end{eqnarray}

Denote
\begin{eqnarray*}
N=\sum_{m=-j}^{j}& &(z_3\frac{m(2Fj+F-j)}{j(j+1)(2j+1)}|m\rangle\langle
m|\\
& &+(z_1+iz_2)\frac{\sqrt{j(j+1)-m(m+1)}(2Fj+F-j)}{2j(j+1)(2j+1)}|m\rangle\langle
m+1|  \\
& &+(z_1-iz_2)\frac{\sqrt{j(j+1)-m(m+1)}(2Fj+F-j)}{2j(j+1)(2j+1)}|m+1\rangle\langle
m|),
\end{eqnarray*}
where $z_1=2(-ty_2+y_1y_3)$, $z_2=2(ty_1+y_2y_3)$, $z_3=t^2+y_3^2-y_1^2-y_2^2$ with $z_1^2+z_2^2+z_3^2=1$. By using the transformation properties of Pauli matrices in \cite{luo,cakmak},
we have the probabilities $p_0=p_1=\frac{1}{2}$, and the corresponding post-measurement states
\begin{eqnarray}
\rho_{0}=(\frac{1}{2j+1}(\sum_{m=-j}^{j}|m\rangle\langle
m|-N)\bigotimes V\Pi_{0}V^\dag
\end{eqnarray}
and
\begin{eqnarray}
\rho_{1}=(\frac{1}{2j+1}\sum_{m=-j}^{j}|m\rangle\langle
m|+N)\bigotimes V\Pi_{1}V^\dag.
\end{eqnarray}

By means of Eqs. (\ref{post}), we have
\begin{eqnarray} \label{r1}
\rho_{A|0}&=&\mbox{tr}_B\rho_{0}\\ \nonumber
&=&\frac{1}{2j+1}(\sum_{m=-j}^{j}|m\rangle\langle
m|-N\\ \nonumber
&=&\frac{1}{2j+1}I_{2j+1}-N,
\end{eqnarray}
and
\begin{eqnarray}\label{r2}
\rho_{A|1}&=&\mbox{tr}_B\rho_{1}\\ \nonumber
&=&\frac{1}{2j+1}\sum_{m=-j}^{j}|m\rangle\langle
m|+N\\ \nonumber
&=&\frac{1}{2j+1}I_{2j+1}+N.
\end{eqnarray}
Furthermore, we have
\begin{eqnarray}
p_{0}\rho_{A|0}+p_{0}\rho_{A|0}=\frac{I_{2j+1}}{2j+1},
\end{eqnarray}
and
\begin{eqnarray}\label{s1}
S(\sum_{i}p_{i}\rho_{A|i})=\log(2j+1).
\end{eqnarray}

According to the properties of tensor product, the eigenvalues of $\rho_{A|0}$ and $\rho_{A|1}$ being equal to the eigenvalues of $\rho_{0}$ and $\rho_{1}$ \cite{wang1} are the same as follow
\begin{eqnarray}
\lambda_n^{\pm}=\frac{1}{2j+1}\pm\frac{j-n}{j(j+1)(2j+1)}\lvert (F(2j+1)-j)\rvert,\label{eigenvalue1}
\end{eqnarray}
where $n=0,\cdots, \lfloor j\rfloor$, and $\lfloor j \rfloor$ denotes the largest integer that is less or equal to $j$.
It is obvious that the eigenvalues are independent of the measurement. Consequently we have
\begin{eqnarray}\label{s2}
\sum_{i}p_{i}S(\rho_{A|i}))&=&\frac{1}{2}S(\rho_{A|0})+\frac{1}{2}S(\rho_{A|1})\nonumber\\
 &=&-\sum_{n=0}^{\lfloor j\rfloor}\lambda_n^{\pm}\log\lambda_n^{\pm}.
\end{eqnarray}
From Eqs. (\ref{fo}), (\ref{s1}) and (\ref{s2}), we have

{\sf [Theorem]} The Holevo quantity of the $SU(2)$-invariant states is given by
\begin{eqnarray}
\chi\{\rho_{AB}|\{\Pi^B_i\}\}=\log(2j+1)+\sum_{n=0}^{\lfloor j\rfloor}\lambda_n^{\pm}\log\lambda_n^{\pm}.
\end{eqnarray}

Fig. 1 shows Holevo quantity as a function of the system parameter $F$ for different $j$. Interestingly, the state $\rho^{ab}$ is separable for $F_s\leq 2j/(2j+1)$ \cite{Schliemann}, which is just twice $F_d=j/(2j+1)$ where the Holevo quantity vanishes. As the dimension of the system increases, the Holevo quantity increases in the region $F<F_d$ and decreases in the region $F>F_d$. The maximum value of the Holevo quantity is attained at $F=1$ for all  dimensional systems. Moreover, as $j$ increases, the maximum value of the Holevo quantity tends to decrease, so does the Holevo quantity, see Fig. 1(a). Furthermore, when $j$ tends to infinite, the Holevo quantity is symmetric around the point $F=1/2$ where Holevo quantity is exactly zero, while its maximum values obtained at $F=0$ and $F=1$ keep unchange, see Fig. 1(b).
\begin{figure}[h]
\raisebox{12em}{(a)}\includegraphics[width=6.25cm]{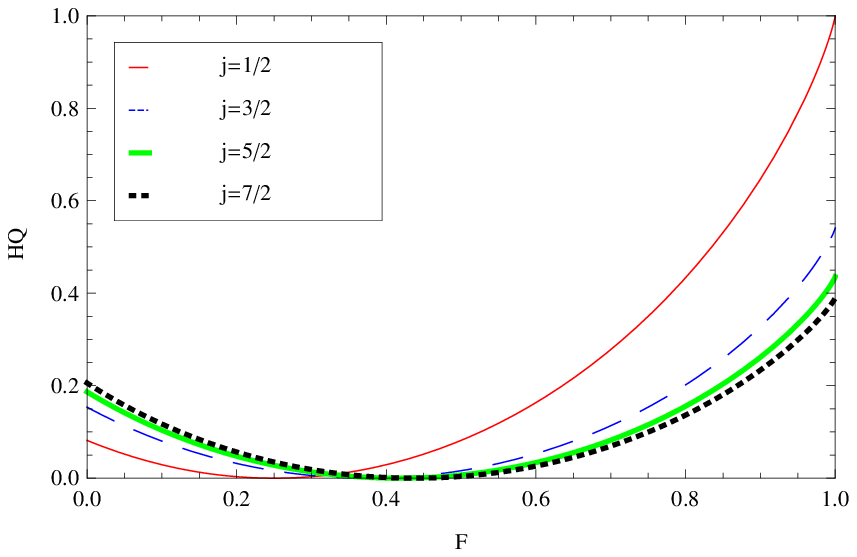}\qquad
\raisebox{12em}{(b)}\includegraphics[width=6.25cm]{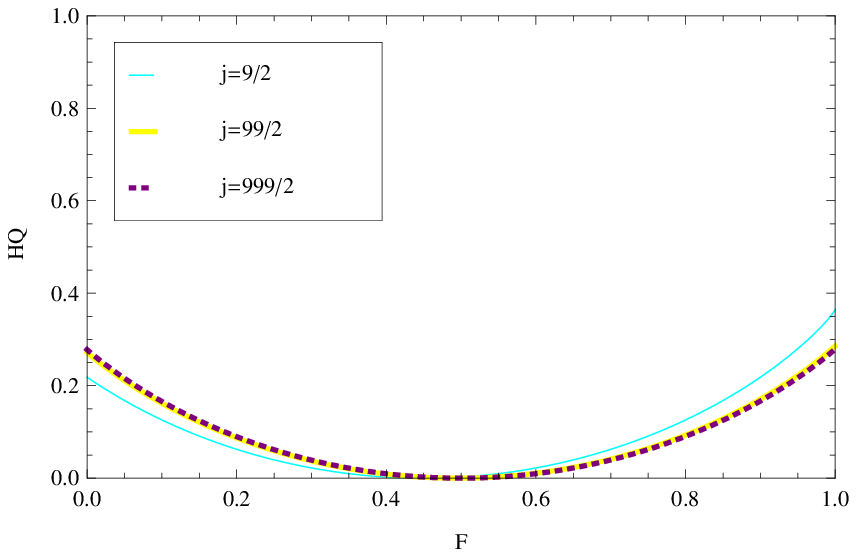}
\label{Fig:1}
\end{figure}
\begin{figure}[h]
\begin{center}
\caption{(Color online) Holevo quantity of $SU(2)$-invariant states as a function of the system parameter $F$ for different $j$.}
\end{center}
\end{figure}

\section{Conclusion}
We have analytically calculated the Holevo quantity of $SU(2)$-invariant states in bipartite spin-$j$ and spin-$\frac{1}{2}$ systems, with the projective measurements performed on the spin-$\frac{1}{2}$ subsystem. It has been shown that the Holevo quantity increases in the region $F<F_d$ and decreases in the region $F>F_d$ when $j$ increases, and the maximum value of the Holevo quantity is attained at $F=1$ for all $j$. The Holevo quantity and the $SU(2)$-invariant states have particular importance in quantum information processing. Our results give explicit relations between the Holevo quantity and the $SU(2)$-invariant states.
\bigskip

\noindent{\bf Acknowledgments}\, \, This work is supported by the National Natural Science Foundation of China (NSFC) under Grant 12065021 and 12075159; Beijing Natural Science Foundation (Grant No. Z190005); Academy for Multidisciplinary Studies, Capital Normal University; Shenzhen Institute for Quantum Science and Engineering, Southern University of Science and Technology (No. SIQSE202001), the Academician Innovation Platform of Hainan Province.

\end{document}